\documentclass[3p,twocolumn]{elsarticle}
\usepackage{epsf}
\newcommand\eq[1] {(\ref{#1})}

\newcommand\labfig[1] {\label{fig:#1}}

\newcommand{\bfm}[1]{\mbox{\boldmath ${#1}$}}
\newcommand{\nonum}{\nonumber \\}
\newcommand{\beqa}{\begin{eqnarray}}
\newcommand{\eeqa}[1]{\label{#1}\end{eqnarray}}
\newcommand{\beq}{\begin{equation}}
\newcommand{\eeq}[1]{\label{#1}\end{equation}}

\newcommand{\Md}{\partial}

\newcommand{\Gve}{\varepsilon}

\newcommand{\Gm}{\mu}

\newcommand{\Go}{\omega}

\newcommand{\BGve}{\bfm\varepsilon}

\newcommand{\BGm}{\bfm\mu}

\newcommand{\BGr}{\bfm\rho}

\newcommand{\BGs}{\bfm\sigma}

\def\Bf{{\bf f}}
\def\Bg{{\bf g}}

\def\Bj{{\bf j}}

\def\Bn{{\bf n}}

\def\Bu{{\bf u}}

\def\Bx{{\bf x}}

\def\BA{{\bf A}}
\def\BB{{\bf B}}
\def\BC{{\bf C}}
\def\BD{{\bf D}}
\def\BE{{\bf E}}
\def\BF{{\bf F}}
\def\BG{{\bf G}}
\def\BH{{\bf H}}
\def\BI{{\bf I}}
\def\BJ{{\bf J}}

\def\BM{{\bf M}}

\def\BS{{\bf S}}

\def\BU{{\bf U}}
\def\BV{{\bf V}}
\def\BW{{\bf W}}
\def\BX{{\bf X}}
\def\BY{{\bf Y}}
\def\BZ{{\bf Z}}

\journal{Physica B}
\begin{document}
\begin{frontmatter}

\title{Hybrid electromagnetic circuits}
\author{Graeme W. Milton}
\address{Department of Mathematics \\ University of Utah \\ Salt Lake City UT 84112 \\ USA}
\author{Pierre Seppecher}
\address{Laboratoire d'Analysis Non Lin\'eaire Appliqu\'ee et Mod\'elisation\\
Universit\'e de Toulon et du Var\\ BP 132-83957 La Garde Cedex\\ France}
\begin{abstract}
Electromagnetic circuits are the electromagnetic analog at fixed frequency of mass-spring networks in
elastodynamics. By interchanging the roles of $\BGve$ and $\BGm$ in electromagnetic circuits
one obtains magnetoelectric circuits. Here we show that by introducing tetrahedral connectors 
having $\BGve=\BGm=0$ one can join electromagnetic and magnetoelectric circuits to obtain hybrid
circuits. Their response is governed by a symmetric matrix with negative semidefinite imaginary 
part. Conversely given any such matrix a recipe is given for constructing a hybrid circuit which
has that matrix as its response matrix. 

\end{abstract}
\end{frontmatter}

At a given fixed frequency $\Go$ Maxwell's equations, which can be written in the form
\beq \frac{\Md}{\Md x_i}\left( C_{ijk\ell}\frac{\Md E_\ell}{\Md x_k}\right)+\{i\Go\Bj\}_j
=-\{\Go^2\BGve\BE\}_j
\eeq{0.1}
where $C_{ijk\ell}=e_{ijm}e_{k\ell n}\{\BGm^{-1}\}_{mn}$ 
[in which $\BE$ is the electric field, $\Bj$ is the free current density
$\BGve$ the electric permittivity tensor,
$\BGm$ the magnetic permeability tensor, and $e_{ijm}=1$ (-1) if $ijm$
is an even (odd) permutation of 123 and is zero otherwise] bear a close
resemblance to the equations of continuum elastodynamics
\beq \frac{\Md}{\Md x_i}\left( C_{ijk\ell}\frac{\Md u_\ell}{\Md x_k}\right)+\{\Bf\}_j
=-\{\Go^2\BGr\Bu\}_j
\eeq{0.3}
[in which $\Bu$ is the displacement field, $\Bf$ is the body-force density 
$\BGr$ is the density tensor, and $\BC$ is now the elasticity tensor]. It is therefore
natural to ask: What is analogous in electrodynamics to a discrete system of springs and masses 
in elastodynamics? The answer is an electromagnetic circuit, introduced by us in \cite{Milton:2009:EC}.
The idea of an electromagnetic circuit generalizes the idea of Engheta, Salandrino, and 
Al\'u \cite{Engheta:2005:CEO} and Engheta \cite{Engheta:2007:CLN} who realized that normal
linear electrical circuits could be approximated, in the quasistatic limit
(which does not imply the frequency is low, but only that the size of the
network is small compared to the wavelength) by a connected network of 
thin cylinders each of material with a suitably scaled value of $\BGve=\Gve\BI$ 
surrounded by a cladding of zero-dielectric material (with $\BGve=0$): a cylinder with
a real positive value of $\Gve$ approximates a capacitor, while a cylinder with a positive imaginary
value of $\Gve$ approximates a resistor, and a cylinder with an almost negative real value of
$\Gve$ approximates an inductor. 

A system of springs and masses can be approximated by massless elastic bars (having
$\BGr=0$ and $\BC$ appropriately scaled; buckling of the bars is ignored as one working within
the framework of linear elasticity) with rigid spherical masses (having $\BC=\infty$ and $\BGr$ 
appropriately scaled) at the junction nodes between bars and 
surrounded by void (having $\BGr=0$ and $\BC=0$). A subset of nodes are chosen to be
terminal nodes at which displacements are prescribed. Similarly
an electromagnetic circuit, as illustrated in Figure 1, 
can be approximated by zero-dielectric diamagnetic thin
triangular plates (of width $h\to 0$ having $\BGve=0$ and $\BGm$ appropriately scaled) with 
zero-magnetic dielectric cylinders (having $\BGm=0$ and $\BGve$ appropriately scaled)
at the junction edges between triangular plates and surrounded by a cladding
(having $\BGve=0$ and $\BGm=\infty$). A subset of edges are chosen to be terminal edges
along which the electric field is prescribed.
As expected from \eq{0.1}-\eq{0.3} $\BGve$ plays the
role of $\BGr$ and $\BGm^{-1}$ plays the role of $\BC$. What is perhaps not quite so expected
is the different geometrical structure of the elements: triangular plates instead of bars and
cylinders instead of spheres. However this is necessitated by the continuity of the fields.
Indeed in the cladding one has $\BH=0$, similarly to the way one has a stress $\BGs=0$
in the void surrounding the elastodynamic network. A bar in such a medium cannot have a constant
non-zero value of $\BH$ inside it, by continuity of the tangential component of $\BH$ across the 
boundary, while a triangular plate in the medium can have a constant value of $\BH$, directed
normal to the surface. When $\Bj=0$ Maxwell's equations remain invariant when the roles of 
($\BE$,$\BD$,$\BGve$,$\Go$) are switched with those of ($\BH$,$\BB$,$\BGm$,-$\Go$). Therefore
for every electromagnetic circuit there is a corresponding magnetoelectric circuit, where the
roles of $\BGve$ and $\BGm$ are interchanged.

\begin{figure}
\vspace{1.2in}
\hspace{0.2in}
{\resizebox{2.0in}{1.0in}
{\includegraphics[0in,0in][5in,2.5in]{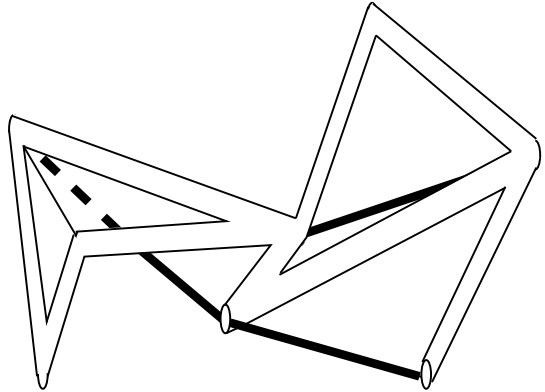}}}
\vspace{0.1in}
\caption{An example of an electromagnetic circuit (EM-circuit) with six zero-dielectric diamagnetic thin
triangular plates (having $\BGve=0$ and $\BGm\ne 0$). It has
small diameter zero-magnetic dielectric cylinders with $\Gve\ne 0$ and $\Gm=0$ 
along a selection of edges including possibly the terminal edges.  
Here there are three terminal edges
marked by thicker solid black lines. 
One internal edge, marked by the thin line, has a cylinder 
with $\Gve=\Gm=0$ attached to it.}
\labfig{4a}
\end{figure}

Of course such extreme values of $\BGve$ and $\BGm$ are difficult to achieve, even with metamaterials 
over a narrow frequency range. So why introduce electromagnetic circuits when they are so difficult
to realize? One answer is that they provide new ways of manipulating electromagnetic fields, that are
relatively easy to analyze. Indeed, the possible responses of electromagnetic circuits in which 
no two terminal edges are connected have been completely characterized \cite{Milton:2009:EC}.
If a desired manipulation is possible with an electromagnetic circuit this should 
motivate the search to achieve a similar manipulation with more realistic materials. Another
answer is more fundamental. Shin, Shen and Fan \cite{Shin:2007:TDE} 
have shown that metamaterials can exhibit macroscopic
electromagnetic behavior which is non-Maxwellian, even though they are governed by Maxwell's equations at the 
microscale. [See also Dubovik, Martsenyuk and Saha \cite{Dubovik:2000:MEE} where other non-Maxwellian macroscopic equations are 
proposed]. So what sort of macroscopic electromagnetic equations can one obtain? The success of 
Camar-Eddine and Seppecher \cite{Camar:2002:CSD,Camar:2003:DCS}
in addressing such questions in the context of three-dimensional (static)
conductivity and elasticity, suggests one should try to characterize the continuum macroscopic behaviors
of electromagnetic circuits and then try to prove that this encompasses all possible macroscopic behaviors.
The continuum limits of electrical circuits can have interesting macroscopic behaviors as discussed
in the book  \cite{Caloz:2006:EM}  and references therein. The continuum limits of electromagnetic 
circuits should have an even richer span of macroscopic behaviors.

When discussing elastic networks it is quite usual to talk about applying a force to a terminal node. 
This could be a concentrated body force, or could be provided by say a medium external to the network 
which we do not need to precisely specify when talking about the applied force. In a similar way in an 
electromagnetic network it is convenient to talk about applying a free electrical current to a terminal
edge. This could be a concentrated current caused by an electrochemical potential, or 
could be an intense $\BH$
field provided by an external medium, acting across the terminal edge. 
To be more precise, at the interface between the external medium 
and the terminal edge we require that $\Bn\cdot\BC\partial \BE/\partial\Bx$ be continuous, which is nothing
more than requiring the tangential component of $\BH$ to be continuous. In elastodynamics the 
quantity $\Bn\cdot\BC\partial \Bu/\partial\Bx$ is called the surface force $\BF$. We adopt a similar
terminology for electrodynamics and call $(i\Go)^{-1}\Bn\cdot\BC\partial \BE/\partial\Bx$ the surface 
free current $\BJ$. (The additional factor of $(i\Go)^{-1}$ is introduced because $i\Go\Bj$ in 
\eq{0.1} plays the role of $\Bf$ in \eq{0.3}.) In a domain $\Omega$ which is divided in 
two subdomains $\Omega_1$, $\Omega_2$ we can say
that $\Omega_2$ exerts on $\Omega_1$ a surface free current $\BJ$ while $\Omega_1$ exerts
on $\Omega_2$ the opposite surface free current $-\BJ$. This formulation does not
mean, in any way, that there exist actual free currents in the material, just like the Newtonian
action-reaction law does not imply the existence of actual surface forces inside the
domain. Similarly in magnetoelectric circuits it is convenient to talk about applying a free magnetic
monopole current to a terminal edge. Again this in no way implies that there exist actual 
magnetic monopole currents, but rather that the equivalent effect is provided by intense 
$\BE$ fields in an external 
medium acting across the terminal edge. But mathematically there is no barrier to thinking of magnetic
monopole currents: in the presence of such currents the equation analogous to \eq{0.1} is
\beq \frac{\Md}{\Md x_i}\left( L_{ijk\ell}\frac{\Md H_\ell}{\Md x_k}\right)-\{i\Go\Bg\}_j
=-\{\Go^2\BGm\BH\}_j
\eeq{0.4}
where $L_{ijk\ell}=e_{ijm}e_{k\ell n}\{\BGve^{-1}\}_{mn}$
and $\Bg$ is the free magnetic monopole current density.

The response of a electromagnetic circuit with $n$ terminal edges is governed by a linear relation 
$i\Go\BJ=\BW\BV$ between the variables $\BJ=(J_1,J_2,\ldots,J_n)$ (not to be confused with 
the $\BJ$ in the previous paragraph) whose components
now represent surface free currents
acting along the terminal edges (the surface free current along terminal edge $i$ is constant along
the edge and directed along the edge, and the complex scalar $J_i$ represents the total current flow
in that direction) and the variables  $\BV=(V_1,V_2,\ldots,V_n)$ which are
the line integrals of the electric field $\BE$ along these edges. As shown in \cite{Milton:2009:EC}, 
the matrix $\BW$ 
is symmetric, with a negative semidefinite imaginary part. Furthermore if the terminal edges are disjoint,
given any matrix $\BW$ with these properties there exists a electromagnetic circuit
(specifically an electromagnetic ladder network, as described in \cite{Milton:2009:EC}) 
which has $\BW$ as its 
response matrix. Similarly, the response of a magnetoelectric circuit with $m$ terminal edges  
is governed by a linear relation $-i\Go\BG=\BY\BU$, or equivalently $\Go\BG=\BY(i\BU)$,
between the variables $\BG=(G_1,G_2,\ldots,G_m)$ 
which represent surface free 
magnetic monopole currents acting on the terminal edges, and the variables  $\BU=(U_1,U_2,\ldots,U_m)$ 
which are the line integrals of the magnetic field $\BH$ along these edges. The matrix $\BY$ 
is symmetric, with a negative semidefinite imaginary part.

Electromagnetic circuits and magnetoelectric circuits seem like completely different animals. Indeed,
an electromagnetic circuit has a cladding with $\BGve=0$ and $\BGm=\infty$, whereas a magnetoelectric
circuit has a cladding with $\BGve=\infty$ and $\BGm=0$. However they can be joined to create hybrid
electromagnetic circuits at the cost of introducing an additional circuit element: a connector 
which joins a terminal edge of an electromagnetic circuit to a terminal edge of a magnetoelectric
circuit. As illustrated in figure 2 the connector is
comprised of a material with $\BGve=\BGm=0$ in the shape of a tetrahedron
with vertices $ABCD$, clad on the top and bottom faces $ACD$ and $BCD$
with a material having $\BGve=0$ and $\BGm=\infty$
(in which $\BH=0$ and $\BD=0$) and on the two side faces $ABC$ and $ABD$
with a material having 
$\BGve=\infty$ and $\BGm=0$ (in which $\BE=0$ and $\BB=0$). The edge $CD$ and 
the edge $AB$ are clipped to expose a surface of width $h\to 0$ with $\BGve=\BGm=0$. The edge $CD$ 
may either be a terminal edge (what we will call an E-terminal), or may be connected to the terminal edge, of width $h$, of an electromagnetic circuit. 
The edge $AB$ may either be a terminal edge (what we will call an H-terminal), 
or may be connected to a terminal edge, of width $h$, of a magnetoelectric circuit. 
\begin{figure}
\vspace{1.0in}
\hspace{0.0in}
{\resizebox{2.0in}{1.0in}
{\includegraphics[0in,0in][3in,1.5in]{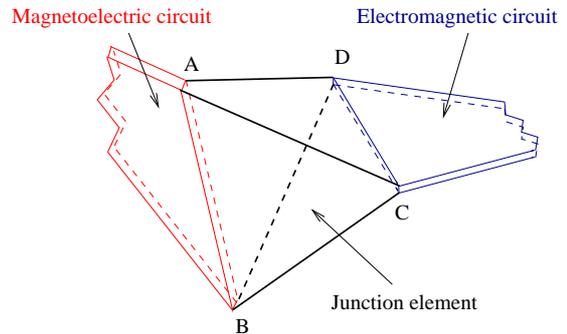}}}
\vspace{0.1in}
\caption{The tetrahedral connector, which allows us to connect a terminal edge in an electromagnetic
circuit (E-terminal) with a terminal edge in a magnetoelectric circuit (H-terminal).}
\labfig{2}
\end{figure}

Inside the tetrahedron $\BGve=\BGm=0$,
and so both $\BE$ and $\BH$ are curl-free. Since $\BH=0$ in the cladding on the top and bottom faces,
and since the tangential component of $\BH$ is continuous across these interfaces, it follows that
the line integral $U_v$ of $\BH$ upwards along the edge $BA$ must equal the line integral of $\BH$ upwards
across the edge $CD$ of width $h$. Similarly, since $\BE=0$ in the cladding on the sides, it follows that
the line integral $V_h$ of $\BE$ forwards along the edge $DC$ must equal the line integral of $\BE$
forwards across the edge $AB$ of width $h$. If $J_h$ denotes the surface free electrical 
current acting along the edge $CD$ in the forward direction and $G_v$ denotes 
the surface free magnetic monopole current acting along the edge $AB$ in the upwards direction,
then $J_h=U_v$ and $G_v=V_h$, or equivalently
\beq \pmatrix{i\Go J_h \cr \Go G_v}=\BM\pmatrix{V_h\cr iU_v}, \eeq{0.5a}
where
\beq
\BM=\pmatrix{0 & \Go \cr \Go & 0}.
\eeq{0.6}

The terminal edges in hybrid electromagnetic circuits come in two varieties: those for which
we can apply a free surface electrical current directed along the edge, which we call E-terminals,
and those for which we can apply (in theory) a free surface magnetic monopole current directed 
along the edge, which we call H-terminals.
The response of general hybrid electromagnetic circuit with $n$ E-terminals
and $m$ H-terminals is governed by a linear
relation $\BA=\BM\BX$ between the variables 
$\BA=(i\Go J_1,i\Go J_2,\ldots,i\Go J_n,\Go G_1,\Go G_2,\ldots,\Go G_m)$ and 
$\BX=(V_1,V_2,\ldots,V_n,i U_1,i U_2,\ldots,i U_m)$,
where the matrix $\BM$ is symmetric, with a negative semidefinite imaginary part. These two
facts are most easily verified if both opposing edges of all tetrahedral connectors are 
included among the terminal edges. Then the matrix $\BM$ takes the form
\beq \BM=\pmatrix{\BW & \BZ \cr \BZ^T & \BY} \eeq{0.7}
where $\BW$ is the response matrix of the (possibly disconnected) part which is an 
electromagnetic circuit, $\BY$ is the response matrix of the (possibly disconnected)
part which is a magnetoelectric circuit and $Z_{ij}$ is $\Go$ if there is a tetrahedral connector
which connects E-terminal $i$ with H-terminal $j$, and is zero otherwise. It then follows, for
example from the arguments in section 5 of \cite{Milton:2009:EC}, that these properties of $\BM$ extend to 
hybrid electromagnetic circuits in which some or all of the tetrahedral connector edges
are not terminal edges. 

Now the response matrix $\BM$ can always be expressed in the form \eq{0.7} if we allow more general (
possibly complex)
matrices $\BZ$. Then by manipulating the relation $\BA=\BM\BX$ we obtain the equivalent relation
\beq \pmatrix{i\Go\BJ \cr i\Go\BU}=\BS\pmatrix{\BV \cr -\BG} \eeq{0.8}
where
\beq \BS=\pmatrix{\BW-\BZ\BY^{-1}\BZ^T & -\Go\BZ\BY^{-1} \cr
                  -\Go\BY^{-1}\BZ^T & -\Go^2\BY^{-1}}
\eeq{0.9}
is symmetric. Conversely it is easy to check that if $\BS$ is symmetric so too is $\BM$. 
Also the fact that
\beqa &~&
-\BX'\cdot\BM''\BX'-\BX''\cdot \BM''\BX'' \nonum
&~&\quad =\BA'\cdot\BX''-\BA''\cdot\BX'
\nonum
&~&\quad =(i\Go\BJ)'\cdot\BV''-(i\Go\BJ)''\cdot\BV' \nonum
&~&\quad\quad
-(i\Go\BU)'\cdot\BG''+(i\Go\BU)''\cdot\BG' \nonum
&~&\quad = -\pmatrix{\BV' \cr -\BG'}\cdot \BS''\pmatrix{\BV' \cr -\BG'}\nonum
&~&\quad\quad
-\pmatrix{\BV'' \cr -\BG''}\cdot \BS''\pmatrix{\BV'' \cr -\BG''}
\eeqa{0.10}
implies $\BM''$ is negative semidefinite if and only if $\BS''$ is negative semidefinite.

Given our hybrid circuit we can attach $m$ tetrahedral connectors to the $m$ H-terminals,
and allow these H-terminals to be internal edges. At the ($n+j$)th E-terminal, which is 
connected to the former $j$th H-terminal we have $J_{n+j}=U_{j}$ and $V_{n+j}=-G_{j}$ where the minus
sign arises, because while $G_{j}$ is the surface free magnetic monopole current acting on
the hybrid circuit at the former $j$th H-terminal, 
$-G_{j}$ is the surface free magnetic monopole current acting on
the tetrahedral connector. According to these relations and \eq{0.8}, $\BS$ will be the response
matrix of this new circuit, i.e. $i\Go\BJ=\BS\BV$. Thus from the hybrid circuit we have obtained
a circuit which responds exactly like a pure electromagnetic circuit. 

We can now establish that any given $(n+m)\times(n+m)$ symmetric matrix $\BS$ with negative 
semidefinite imaginary part can be realized by a hybrid circuit with $n$ E-terminals 
and $m$ H-terminals which have no vertex in common. We first
construct the $n+m$ E-terminal electromagnetic ladder network which has $\BS$ as its response
matrix. Then to the edges $n+1,n+2,\ldots, n+m$ we attach tetrahedral connectors to convert these
E-terminals into H-terminals (taking these former E-terminals to be internal edges in the new circuit).
This leaves the response matrix $\BS$ unchanged, which now governs the response \eq{0.8} of our
new hybrid circuit. The proof is complete. Note that in the ladder network (and more generally
in other electromagnetic circuits), the internal zero-magnetic dielectric cylinders which join the terminal
edge will generally carry some displacement current $\BD$. When we join a tetrahedral connector to
the terminal edge this displacement current will flow into the side cladding which has
$\BGve=\infty$ and $\BGm=0$ and can support a non-zero value of $\BD$.

In summary, although hybrid electromagnetic circuits seem to be vastly more general than
pure electromagnetic or magnetoelectric circuits, they are in a sense (modulo the addition
of tetrahedral connectors to the terminal edges) all equivalent. 

\section*{Acknowledgements}
G.W.M. is grateful for support from the 
National Science Foundation through grant DMS-0707978.

\bibliography{/u/ma/milton/tcbook,/u/ma/milton/newref}

\end{document}